\documentclass{PoS}
\usepackage{graphicx}
\usepackage{subfigure}
\usepackage{gensymb}
\usepackage{siunitx}

\title{Test results of a prototype device to calibrate the Large Size Telescope camera proposed for the Cherenkov Telescope Array}

\ShortTitle{Test results of a prototype device to calibrate the LST camera of the CTA}

\author{{M. Palatiello}\thanks{michele.palatiello@ts.infn.it}\,\,$^1$, D. Cauz\,$^2$, F. De Persio\,$^3$, F. Ferrarotto\,$^3$, G. Pauletta\,$^2$ and M. Iori\,$^3$ \\
        $^1$ INFN Trieste and University of Trieste, Italy \\
        $^2$ INFN Trieste and University of Udine, Italy \\
        $^3$ INFN Sezione di Roma1 and Rome Sapienza, Italy}
\author{P.  Majumdar\,$^4$, A. Chatterjee\,$^5$, V. Chitnis\,$^5$, B.B. Singh\,$^5$, K. Gothe\,$^5$  , M. Manoranjan\,$^5$  \\
        $^4$ Saha Institute of Nuclear Physics, Calcutta, India\\
        $^5$ Tata Institute of Fundamental Research, Mumbai, India}
\author{for the CTA Consortium\footnote{https://www.cta-observatory.org/about/cta-consortium/}}

\abstract{A Large Size air Cherenkov Telescope (LST) prototype, proposed for the Cherenkov Telescope Array (CTA), is under construction at the Canary Island of La Palma (Spain) this year. The LST camera, which comprises an array of about 500 photomultipliers (PMTs), requires a precise and regular calibration over a large dynamic range, up to $10^3$ photo-electrons (pe's), for each PMT. We present a system built to provide the optical calibration of the camera consisting of a pulsed laser (355 nm wavelength, 400 ps pulse width), a set of filters to guarantee a large dynamic range of photons on the sensors, and a diffusing sphere to uniformly spread the laser light, with flat fielding within 3\%, over the camera focal plane 28 m away. The prototype of the system developed at INFN is hermetically closed and filled with dry air to make the system completely isolated from the external environment. In the paper we present the results of the tests for the evaluation of the photon density at the camera plane, the system isolation from the environment, and the shape of the signal as detected by the PMTs. The description of the communication of the system with the rest of detector is also given.}

\FullConference{35th International Cosmic Ray Conference ICRC217\\
		10-20 July, 2017\\
		Bexco, Busan, Korea}

\begin{document}

\section{Introduction}
A periodic calibration of the LST camera is crucial for an accurate reconstruction of the Cherenkov event. The Calibration System generates triggers, which are synchronous to the light pulses because they must be distinguished from real physics events. In addition, it produces random triggers that are used to study the pedestal level \cite{doc2}.
 For this purpose we have built a INFN \textit{Camera Calibration Box} (hereafter CaliBox), located at the center of the parabolic \SI{23}{\metre} diameter mirror structure of the LST telescope, which will flash the \SI{28}{\metre} away camera with a \SI{1}{\micro\joule} UV laser source \cite{doc2,doc1,laser}. By this procedure  the ratio between the number of digital counts recorded by the Readout System and the pe's is determined.
The laser generates \SI{400}{\pico\second} UV pulses at a rate between \SI{1}{\hertz} and \SI{2000}{\hertz} and the combination of a double set of neutral density filters makes it possible to achieve the required dynamic range from 10 to 80000 pe's at the PMTs as is explicitly required by the Technical Design Report LST for the CTA Consortium \cite{doc2}. The device is completely managed and controlled by an Open Platform Communication Server (OPC-S) completely installed on a ODROID-C1 board \cite{mos}.
These proceedings are outlined as following: in section 2 the frame, electronics and optics of CaliBox are discussed. Section 3 discusses the evaluation of the number of photons at the camera focal plane, section 4 presents the results of the flat fielding uniformity measurements. 

\section{The Frame of the CaliBox}
The casing of the INFN CaliBox, as shown in Figure 1, consists of a thermal radiator aluminum plate to dissipate the heat generated by all the components ($\sim$\SI{100}{\watt}). The two boxes, optically connected, are fixed on the plate, one contains the laser and filter wheels \cite{wheel} the other contains a 1 inch diffuser and two different photon detectors. These two boxes are closed hermetically by o-rings hence the optical system is isolated from the external environment. A \SI{100}{\watt} power supply and the ODROID-C1 computer board are placed outside of the boxes and all is contained in an aluminum water proof shell. A sensor inside the hermetic box and another one located close to the electronics continuously monitor the temperature and the relative humidity (RH).
\\
 \begin{figure}[htbp]
        \centering%
        \subfigure[]%
          {\includegraphics[width=70mm]{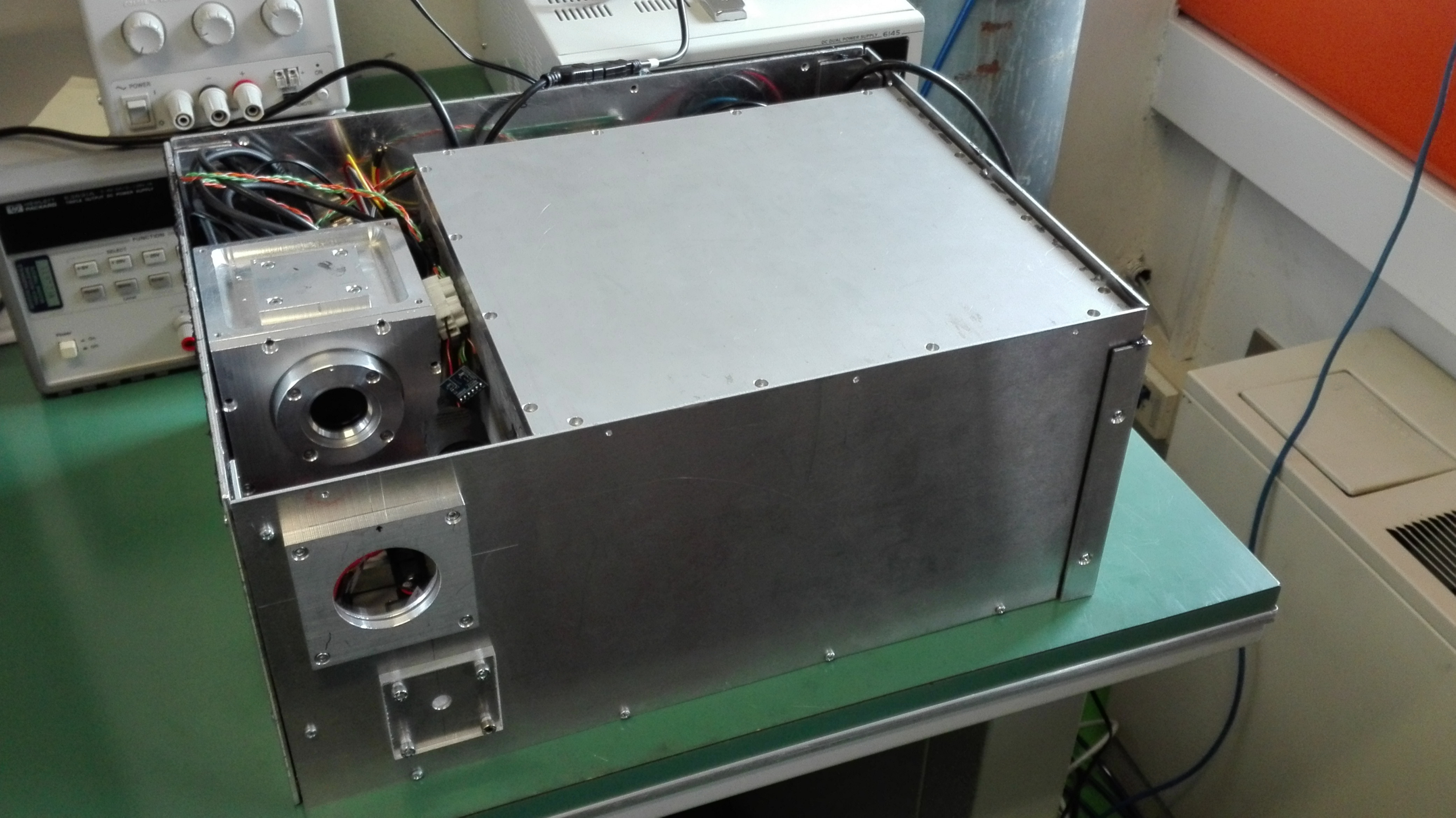}}\qquad
        \subfigure[]%
          {\includegraphics[width=70mm]{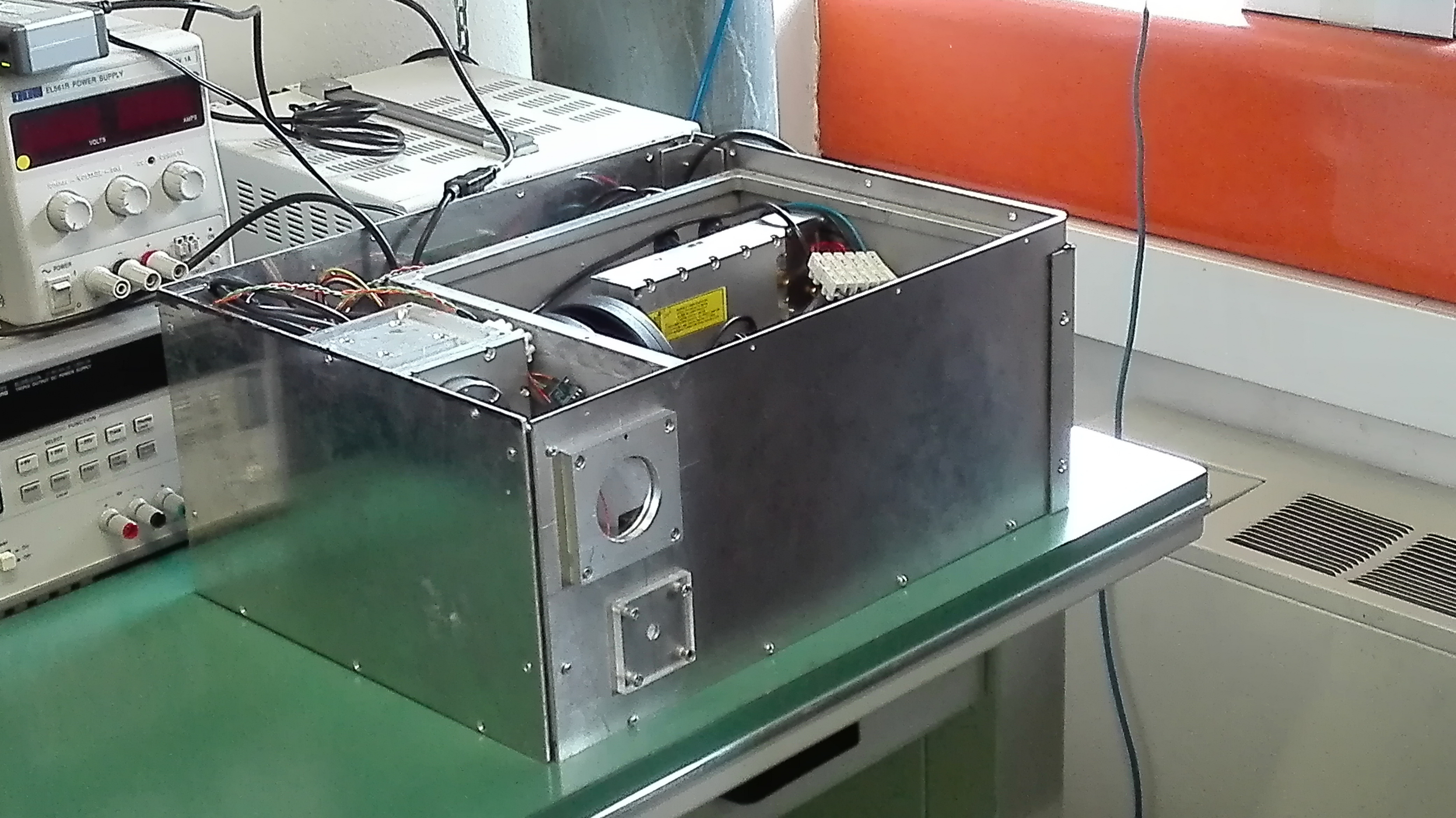}}
        \caption{(a) The CaliBox without the lid; on the left is visible the diffuser box with the exit window; on the left of the front side the UV beam and the pointer laser exits are visible. (b) The open boxes showing the internal instrumentation.\label{fig:stofigure}}
\end{figure}

The monitoring of the dynamic range of photons is provided by two different sensors (Photodiode Hamamatsu 3590-18 (PD) and a Sensl 3x3 \si{\square\milli\metre} Silicon-Photomultiplier (SiPM)), covering the large photon intensity. These sensors are placed near one of the exit ports of the diffuser and read by a 10 bit ADC allowing to evaluate the number of pe's sent to the camera. The CaliBox has been designed to have good dissipation of the heat produced by the electronics and the laser and is supposed to operate in high RH environment \cite{laser}. 
For this reason, the box containing the laser, wheels and diffuser is hermetically closed by covers with o-rings and is filled with dry air at the pressure of 1 atm and the whole shell is IP67 certified. This ensures a stable and low RH (20\%) inside the box, while the thermal conductivity will avoid the water vapor to condensate. To verify the performances of the CaliBox we have performed several tests in a climatic chamber at the University of Udine measuring the RH, as a function of temperature and time by using a weather-board connected to a ODROID-C1. To test the isolation from the environment, we placed the hermetically closed CaliBox inside the climatic chamber, where the RH was raised to the value of 95\% and the CaliBox was filled with dry air until the internal RH reached the value of 20\%. The CaliBox RH was monitored for 12 hours and resulted stable at $20\% \pm 2\%$. Having a source of heat inside the box and filling with dry air we are constraining the dew point, i.e. with a RH of 20\% the dew point temperature ranges between \ang{0}\,C and \ang{10}\,C for a temperature inside the box between \ang{25}\,C and \ang{40}\,C. 
\section{Calibration method and Evaluation of photon density and uniformity}
To calibrate the camera we need a signal with the same characteristics as the pulse generated by the Cherenkov light produced by the shower over the single PMT that results in a light pulse with a FWHM of $2.85\pm0.03\,$\si{\nano\second} (using MAGIC data, Figure 2 (a)).
\\
 \begin{figure}[htbp]
        \centering%
        \subfigure[]%
          {\includegraphics[width=65mm]{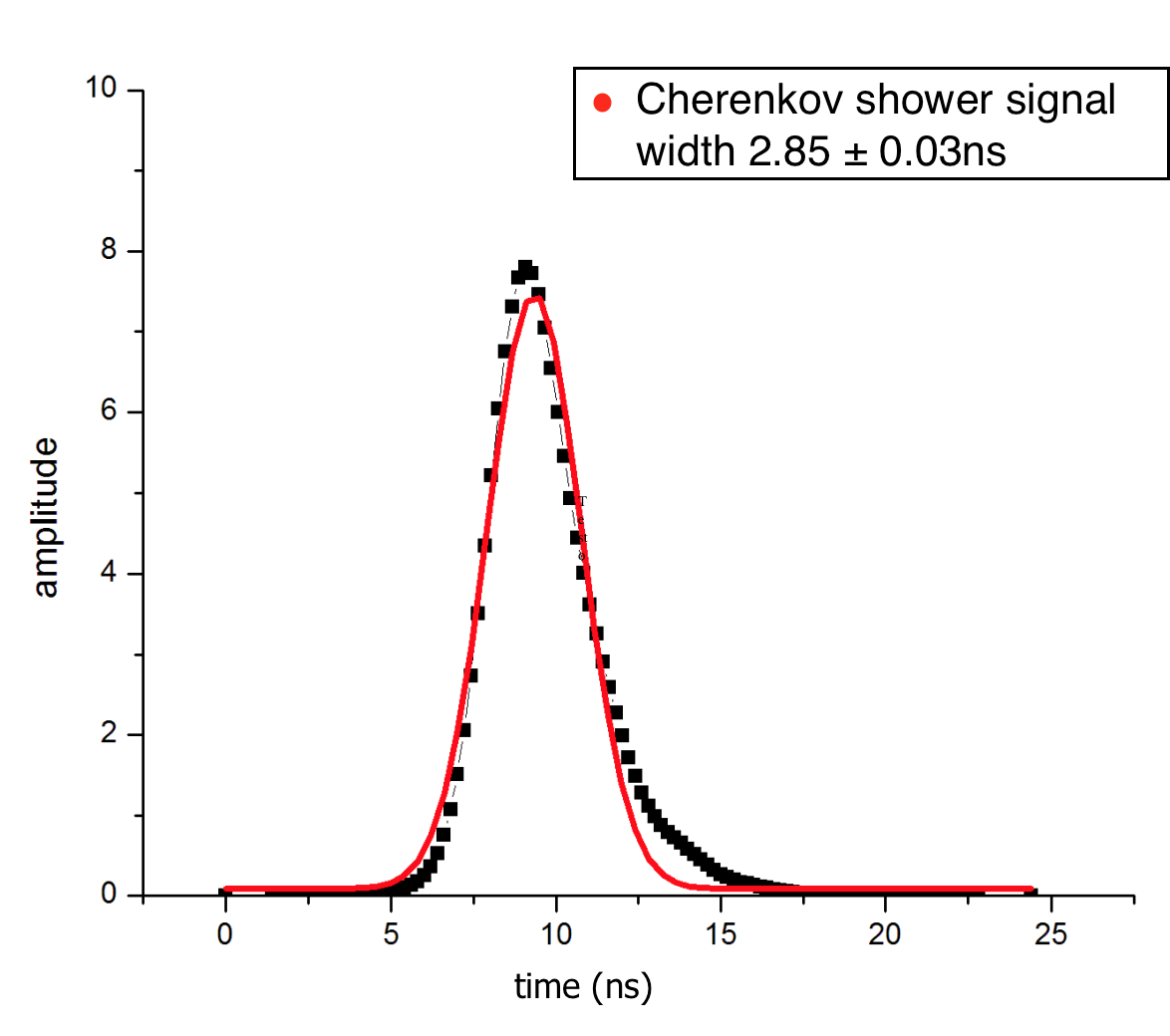}}\qquad
        \subfigure[]%
          {\includegraphics[width=78mm]{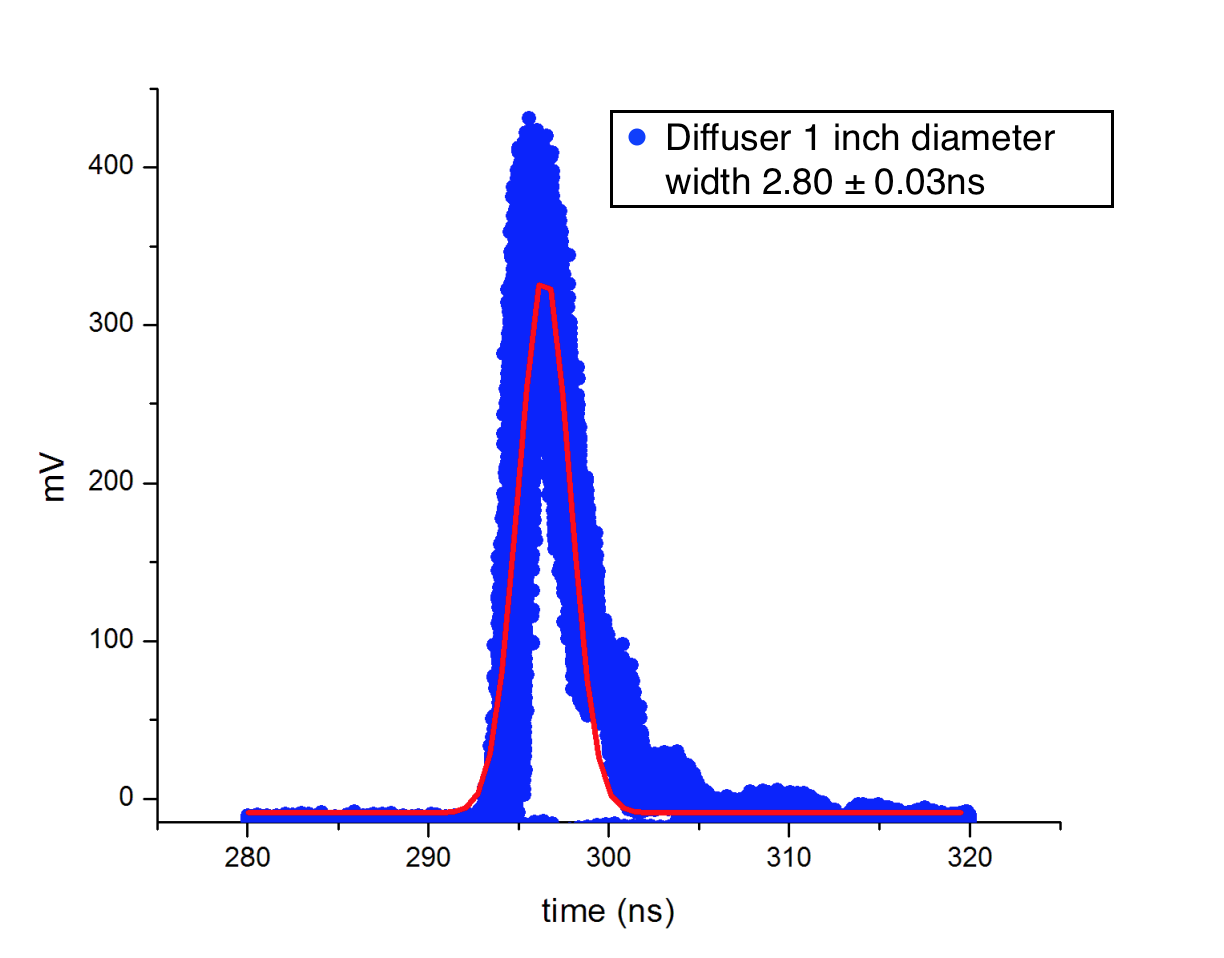}}
        \caption{(a) Cherenkov signal as seen by the MAGIC camera. The Amplitude is in mV and the time in \si{\nano\second}. The width, obtained by a Gaussian fit of the signal, is $2.85\pm0.03\,$\si{\nano\second}. (b) Pulse at exit of the 1 inch diffuser. The solid red line is the results of Gaussian fit that gives a width of $2.80\pm0.03\,$\si{\nano\second}.\label{fig:soe}}
\end{figure}

The width of the pulse generated by our system depends on the diameter of the diffuser sphere and it results to have a pulse shape consistent with the Cherenkov signal with a FWHM of $2.80\pm0.03\,$\si{\nano\second} (as shown in Figure 2 (b)). The PMTs calibration requires a high photon dynamic range and uniformity at the camera plane.
We have performed a test to evaluate the photon density using a statistical method of pe's calibration. Considering that the photons released by the laser at each shot follow a Gaussian distribution and that the probability of detection of photons reaching the sensor can be described by a binomial distribution we obtain a normal distribution with a $\sigma^{2}$ proportional to the average number of photons and a constant related to the photon dispersion at source. Hence if we repeat the measurement for different filters and perform a quadratic fit of the data, the linear term of the fit allows us to determine the number of photons reaching the camera \cite{statist}. The measurements were done using filters with optical density (OD) from 2 to 3 and the signal shape was detected by the SiPM (overvoltage of \SI{500}{\milli\volt} above \SI{27.3}{\volt} breakdown voltage with a constant temperature of \ang{25}\,C) \SI{5}{\metre} distant from the CaliBox using a DRS4 \cite{drs4} to digitize the signal. Figure 3 (a) shows the dependence of $\sigma^{2}$ of the integrated signal value on different filters resulting in a fit value of 15 nVs/pe.
Taking into account the measured transmission factor and using the quantum efficiency of the sensor we have evaluated the number of ph/\si{\square\centi\metre} at the distance of \SI{5}{\metre} which extrapolating to \SI{28}{\metre}, results to be $6.4 \pm 0.2 \times 10^{4}$ ph/\si{\square\centi\metre}.
\\
 \begin{figure}[htbp]
        \centering%
        \subfigure[]%
          {\includegraphics[width=68.2mm]{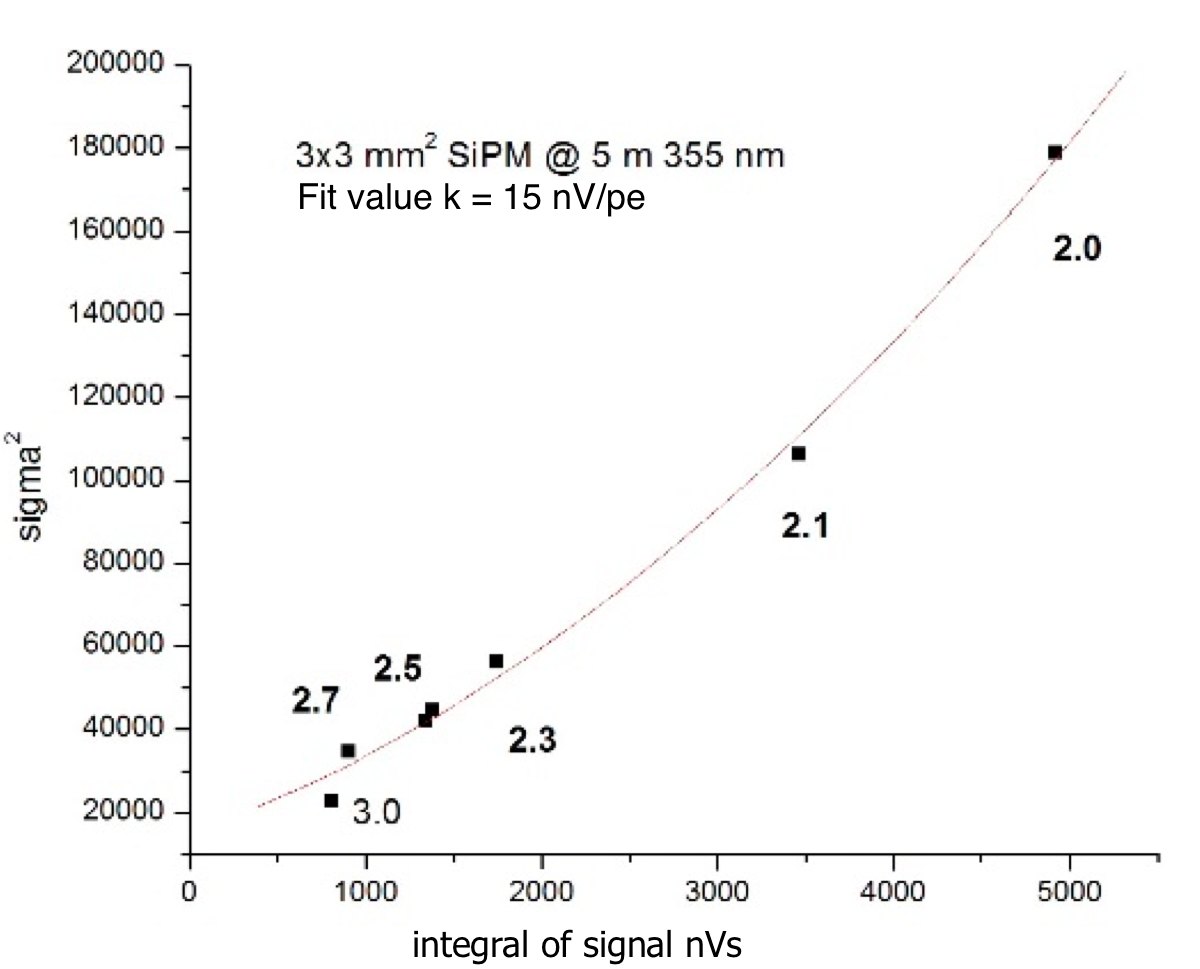}}\qquad
        \subfigure[]%
          {\includegraphics[width=74.8mm]{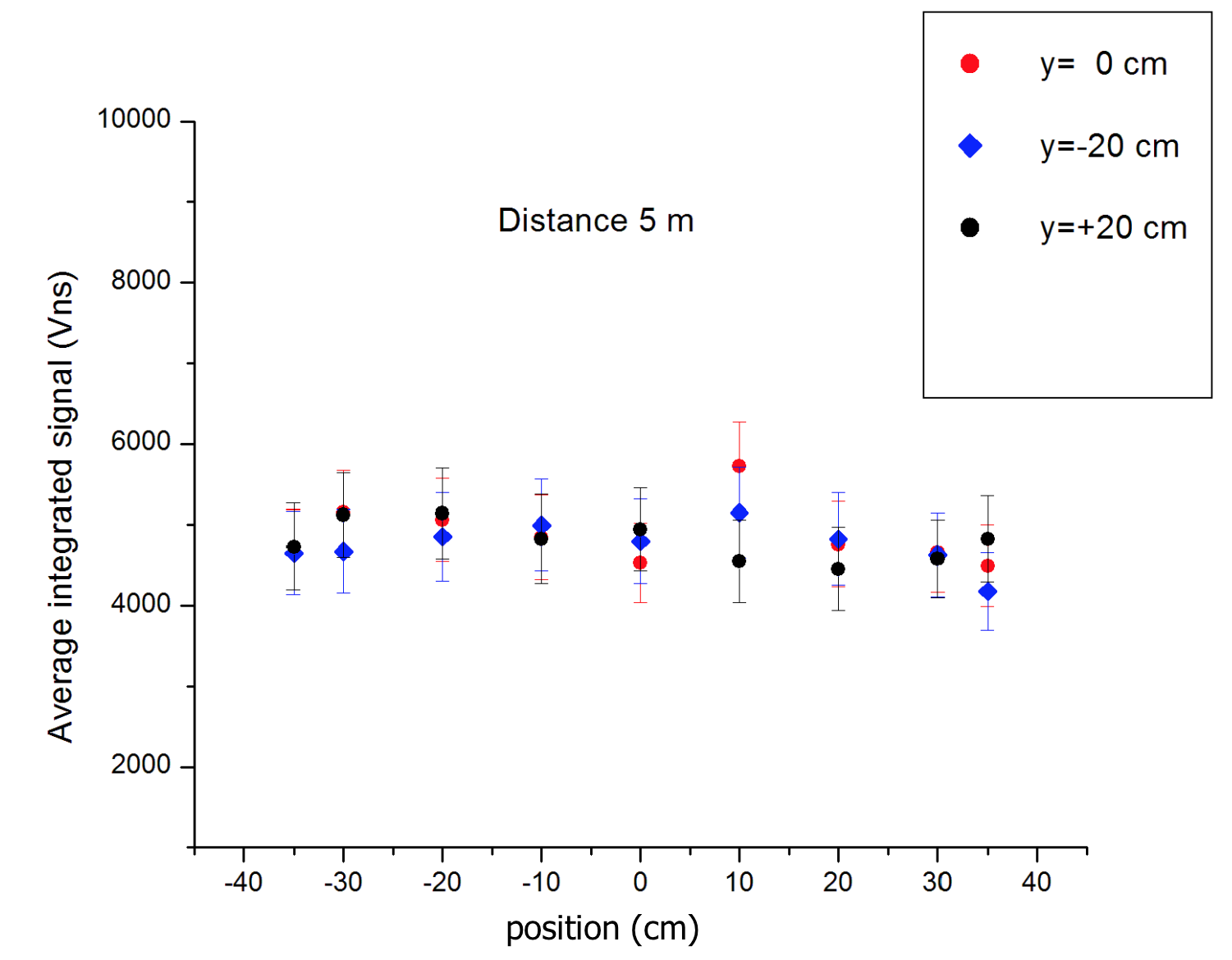}}
        \caption{(a) The $\sigma^{2}$ of the integrated signal value versus signal integral mean for different OD. The OD values are printed along the curve and they range from 2 to 3. The curve is the best fit result and the values of the linear term of the quadratic function is 15 nVs/pe. At OD = 2.5 we took two sets of measures to verify our measurement fluctuation. (b) The SiPM average integrated signal vs. position on $x$ (range -\SI{40}{\centi\meter} to +\SI{40}{\centi\meter} from the center of the beam spot) for the three $y$ vertical positions (blue at -\SI{20}{\centi\meter}, red at \SI{0}{\centi\meter} and black at +\SI{20}{\centi\meter} from the center of the beam spot). The SiPM was located 5 m away from the CaliBox operating at \ang{25}\,C.\label{fig:sottof}}
\end{figure}
 The uniformity of photons has also been measured at 5 m from the CaliBox using a SiPM successively placed on 27 points of the plane and lying within a cone view angle of $5\degree$ from the laser beam. To evaluate the uniformity of the diffuse light at the camera plane we performed a linear fit obtaining a constant value $a_{f}$. Evaluating the average of each set of measurements at the three different $y_{n}$ positions, $<a_{y_{n}}>$ as shown in Figure 3 (b), the average ratio of the absolute $a_{f}$ and $<a_{y_{n}}>$ difference, for the three sets (red, blue and black $y$ positions in Figure 3 (b)), divided by the $a_{f}$ results to be about 3\%.

\section{Remote control of the INFN CaliBox}
The single board class computer ODROID-C1 is equipped with a LST Ubuntu Linux operative system that allows to run all the programs and plugins for the correct use and control of all the device instrumentation.
The central control system is completely remotely managed (by Secure SHell protocol) by an OPC-Server based on a Media Object Server Communications Protocol (MOS) \cite{mos} and implemented in the ODROID-C1. The MOS software program converts the commands into the OPC-Server plugin protocol. We have adapted a OPC-Client \cite{client} that uses the OPC-Server to get data from or send commands to the CaliBox to set the laser parameters, change the filter combination, start and stop the red pointing laser and get the value of T and RH inside the box.

\section{Conclusions}
The CTA is a proposed new generation of gamma ray observatory with high sensitivity. To achieve precise measurements of astronomical sources we need continuous calibration of the camera. In the tests of the INFN CaliBox prototype we have measured a width of the pulse out of a 1 inch diffuser to be $2.80\pm0.03$\,\si{\nano\second} consistent with the Cherenkov signal read by the camera, a photon density at the camera plane of order $10^{4}$ ph/\si{\square\centi\metre}. An accuracy of about 3\% in uniformity of light intensity has been measured within $\pm$\,\ang{5}  along the laser beam at a distance of 5 m from the INFN CaliBox. Tests on isolation of the INFN CaliBox in a climatic chamber show the system is free of water vapor condensation and when filled with dry air the RH is stable at a value of 20\%. A completely remote OPC-Server control of the INFN CaliBox has been adapted in the ODROID platform to remotely control all the devices.
\\

\section*{\centering ACKNOWLEDGMENTS}
Special thanks are due to Dr. Jean-Luc Panazol, Dr. Thierry Le-Flour  and Dr. Julie Prast of Laboratoire d'Annecy-le-Vieux de Physique des Particules (Université de Savoie, France) for the support and willingness in helping us in adapting the OPC Server \cite{mos} for our devices.
We would also like to thank Dr. L. Recchia, G. Chiodi, R. Lunadei of INFN-RM1 Electronics Laboratory (LABE) and M. Iannone and M. Ciaccafava and Dr. A. Zullo of INFN-RM1-Engineering Design Office and A. Sabatini of University and INFN of Udine Mechanical Workshop.
We gratefully acknowledge financial support from the agencies and organizations listed here: http://www.cta-observatory.org/consortium\_acknowledgments


\begin{thebibliography}{99}
\bibitem{doc2}
Technical Design Report LST https://www.cta-observatory.org/
\bibitem{doc1}
M. Iori et al., PoS ICRC2015 (2016) 954 CTA Consortium
\bibitem{laser}
 https://www.teemphotonics.com/laser/
 \bibitem{mos}
Media Object Server Communications Protocol (MOS) provided by the Laboratoire d'Annecy-le-Vieux de Physique des Particules (LAPP)
\bibitem{wheel}
https://www.thorlabs.de/
\bibitem{statist}
A.T. Fienberg et al., Studies of an array of PbF2  Cherenkov crystals with large-area SiPM readout, Nuclear Instruments and Methods in Physics Research A783 (2015) 12-21
\bibitem{drs4}
https://www.psi.ch/drs/DocumentationEN/manual\_rev50.pdf
\bibitem{client}
https://www.unified-automation.com/products/development-tools/uaexpert.html

\end{thebibliography}
\end{document}